\documentclass[letterpaper]{article} % DO NOT CHANGE THIS
\usepackage[]{aaai24}  % DO NOT CHANGE THIS
\usepackage{times}  % DO NOT CHANGE THIS
\usepackage{helvet}  % DO NOT CHANGE THIS
\usepackage{courier}  % DO NOT CHANGE THIS
\usepackage[hyphens]{url}  % DO NOT CHANGE THIS
\usepackage{graphicx} % DO NOT CHANGE THIS
\usepackage{natbib}  % DO NOT CHANGE THIS AND DO NOT ADD ANY OPTIONS TO IT
\usepackage{caption} % DO NOT CHANGE THIS AND DO NOT ADD ANY OPTIONS TO IT
\usepackage{algorithm}
\usepackage{algorithmic}
\usepackage{amsmath,amssymb,amsfonts}
\usepackage{mathtools}
\usepackage{graphicx}
\usepackage{xspace}
\usepackage{soul}
\usepackage[flushleft]{threeparttable}
\usepackage{multirow}
\usepackage{booktabs}
\usepackage{color}
\usepackage{subcaption}

\newcommand{\method}{\mbox{$\mathop{\mathtt{SSNA}}\limits$}\xspace}
\newcommand{\TT}{\mbox{$\mathop{\mathtt{TT}}\limits$}\xspace}
\newcommand{\FT}{\mbox{$\mathop{\mathtt{MLT}}\limits$}\xspace}
\newcommand{\AM}{\mbox{$\mathop{\mathtt{AM}}\limits$}\xspace}
\newcommand{\CNN}{\mbox{$\mathop{\mathtt{CNN}}\limits$}\xspace}
\newcommand{\GRU}{\mbox{$\mathop{\mathtt{GRU}}\limits$}\xspace}
\newcommand{\GRUs}{\mbox{$\mathop{\mathtt{GRUs}}\limits$}\xspace}

\newcommand{\LoRA}{\mbox{$\mathop{\mathtt{LoRA}}\limits$}\xspace}
\newcommand{\ADALoRA}{\mbox{$\mathop{\mathtt{ADA}}\limits$}\xspace}
\newcommand{\SASRec}{\mbox{$\mathop{\mathtt{SASRec}}\limits$}\xspace}
\newcommand{\Full}{\mbox{$\mathop{\mathtt{Full}}\limits$}\xspace}
\newcommand{\TTMLP}{\mbox{$\mathop{\mathtt{TMLP}}\limits$}\xspace}
\newcommand{\TTMoE}{\mbox{$\mathop{\mathtt{TMoE}}\limits$}\xspace}
\newcommand{\BitFit}{\mbox{$\mathop{\mathtt{BitFit}}\limits$}\xspace}

\newcommand{\MEAN}{\mbox{$\mathop{\mathtt{MEAN}}\limits$}\xspace}
\newcommand{\WEIGHTED}{\mbox{$\mathop{\mathtt{WEIGHTED}}\limits$}\xspace}
\newcommand{\BERT}{\mbox{$\mathop{\mathtt{BERT}}\limits$}\xspace}
\newcommand{\RoBERTa}{\mbox{$\mathop{\mathtt{RoBERTa}}\limits$}\xspace}
\newcommand{\BERTMedium}{\mbox{$\mathop{\mathtt{BERTMedium}}\limits$}\xspace}
\newcommand{\DistillBERT}{\mbox{$\mathop{\mathtt{DistilBERT}}\limits$}\xspace}
\newcommand{\DistillRoBERTa}{\mbox{$\mathop{\mathtt{DistilRoBERTa}}\limits$}\xspace}

\newcommand{\SCI}{\mbox{$\mathop{\mathtt{Sci}}\limits$}\xspace}
\newcommand{\Pantry}{\mbox{$\mathop{\mathtt{Pantry}}\limits$}\xspace}
\newcommand{\INS}{\mbox{$\mathop{\mathtt{Ins}}\limits$}\xspace}
\newcommand{\Tools}{\mbox{$\mathop{\mathtt{Tools}}\limits$}\xspace}
\newcommand{\Toys}{\mbox{$\mathop{\mathtt{Toys}}\limits$}\xspace}

\newcommand{\RK}{\mbox{$\mathop{\mathtt{R@k}}\limits$}\xspace}
\newcommand{\NK}{\mbox{$\mathop{\mathtt{N@k}}\limits$}\xspace}
\newcommand{\RT}{\mbox{$\mathop{\mathtt{R@10}}\limits$}\xspace}
\newcommand{\NT}{\mbox{$\mathop{\mathtt{N@10}}\limits$}\xspace}
\newcommand{\RF}{\mbox{$\mathop{\mathtt{R@50}}\limits$}\xspace}
\newcommand{\NF}{\mbox{$\mathop{\mathtt{N@50}}\limits$}\xspace}
\newcommand{\etal}{\textit{et al}.}

\usepackage{newfloat}
\usepackage{listings}
\DeclareCaptionStyle{ruled}{labelfont=normalfont,labelsep=colon,strut=off} % DO NOT CHANGE THIS
\floatstyle{ruled}
\newfloat{listing}{tb}{lst}{}
\floatname{listing}{Listing}
\title{Towards Efficient and Effective Adaptation of Large Language Models for Sequential Recommendation}
\author{
    Bo Peng, \textsuperscript{\rm 1}
    Ben Burns, \textsuperscript{\rm 1}
    Ziqi Chen, \textsuperscript{\rm 1}
    Srinivasan Parthasarathy, \textsuperscript{\rm 1, 3}
    Xia Ning \textsuperscript{\rm 1, 2, 3}
}
\affiliations{
    \textsuperscript{\rm 1} 
    Department of Computer Science and Engineering, The Ohio State University, Columbus\\
    \textsuperscript{\rm 2}
    Department of Biomedical Informatics, The Ohio State University, Columbus\\
    \textsuperscript{\rm 3}
    Translational Data Analytics Institute, The Ohio State University, Columbus\\
    \{peng.707, burns.1241, chen.8484\}@buckeyemail.osu.edu, 
    srini@cse.ohio-state.edu, 
    ning.104@osu.edu
}

\usepackage{bibentry}
\begin{document}

\maketitle

\begin{abstract}

In recent years, with large language models (LLMs) achieving state-of-the-art performance in context understanding, increasing efforts have been dedicated to developing LLM-enhanced sequential recommendation (SR) methods.
Considering that most existing LLMs are not specifically optimized for recommendation tasks, adapting them for SR becomes a critical step in LLM-enhanced SR methods.
Though numerous adaptation methods have been developed, it still remains a significant challenge to adapt LLMs for SR both efficiently and effectively.
To address this challenge, in this paper, we introduce a novel side sequential network adaptation method, denoted as \method, for LLM-enhanced SR.
\method features three key designs to allow both efficient and effective LLM adaptation.
First, \method learns adapters separate from LLMs, while fixing all the pre-trained parameters within LLMs to allow efficient adaptation.
In addition, \method adapts the top-$a$ layers of LLMs jointly, and integrates adapters sequentially for enhanced effectiveness (i.e., recommendation performance).
We compare \method against five state-of-the-art baseline methods on five benchmark datasets using three LLMs.
The experimental results demonstrate that \method significantly outperforms all the baseline methods in terms of recommendation performance, and achieves substantial improvement over the best-performing baseline methods at both run-time and memory efficiency during training.
Our analysis shows the effectiveness of integrating adapters in a sequential manner.
Our parameter study demonstrates the effectiveness of jointly adapting the top-$a$ layers of LLMs.

\end{abstract}

%%%%%%%%%%%%%%%%%%%%%%%%%%%%%%%%%%%%%%%%%%%%%%%%%
\section*{Introduction}
\label{sec:introduction}
%%%%%%%%%%%%%%%%%%%%%%%%%%%%%%%%%%%%%%%%%%%%%%%%%

Sequential recommendation (SR) aims to predict and recommend the next item of users' interest using their historical interactions.
It has been
drawing increasing attention from the research community due
to its wide applications in online retail, video streaming and tourism planning, etc.
Recently, as large language models (LLMs) have demonstrated state-of-the-art performance in context understanding~\cite{OpenAI2023GPT4TR}, numerous efforts have been dedicated to developing LLM-enhanced SR methods~\cite{yuan2023go,hou2022towards}.
The key idea behind these methods is to leverage LLMs to generate expressive item embeddings from item texts (e.g., titles) for better recommendation.
Given that existing LLMs are generally recommendation-independent, 
adapting LLMs for recommendation tasks becomes an important step in LLM-enhanced SR methods.
%a critical challenge in developing LLM-enhanced SR methods is to both efficiently
%~\footnote{In this paper, we focus on the computation and memory efficiency of methods.} 
%and effectively adapt LLMs for recommendation tasks. 
%
%Considering that existing LLMs are generally recommendation-independent, 
%
%adapting them to SR becomes a critical step in enabling effective recommendation.
%

\begin{figure}[!t]
       \centering
       \footnotesize
        \begin{minipage}{\linewidth}
               \begin{subfigure}{0.48\linewidth}
                    \centering
                    \includegraphics[width=0.6\linewidth]{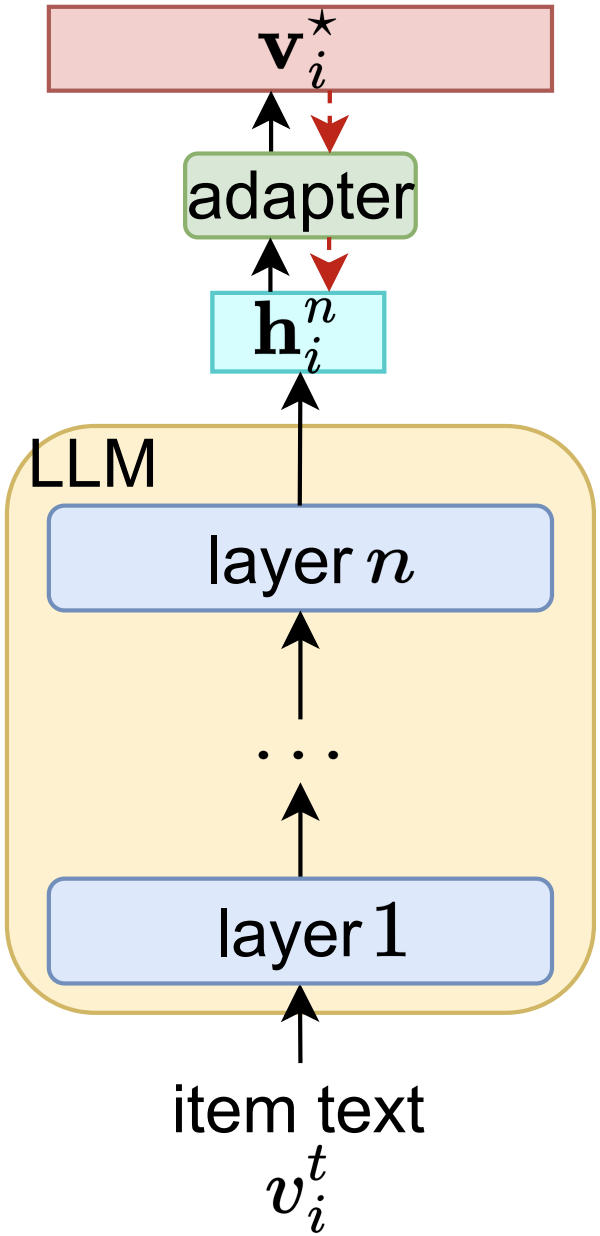}
                    \caption{\TT}
                    \label{fig:tlt}
                \end{subfigure}
                \begin{subfigure}{0.48\linewidth}
                    \centering
                    \includegraphics[width=0.6\linewidth]{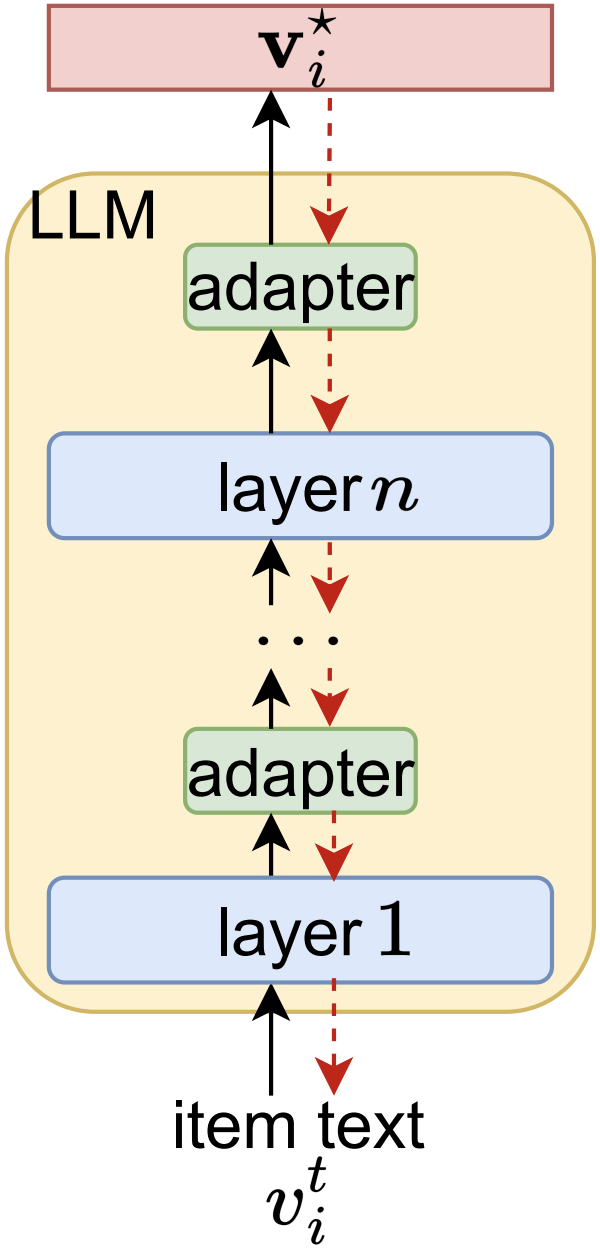}
                    \caption{\FT}
                    \label{fig:mt}
                \end{subfigure}
    \end{minipage}
\caption{
Illustration of \TT (left) and \FT (right).
In this figure, $\mathbf{h}^n_i$ is the embedding generated from the LLM using $v_i^t$; and $\mathbf{v}_i^\star$ is the adapted item embedding.
This figure shares the same legend as Figure~\ref{fig:architecture}.
%
%Note that \TT could avoid forward and backward propagation across LLMs in training if pre-calculated $\mathbf{h}_i^n$ for all the items.
%\xia{fonts are too small in the figure to visualize.}
}
\label{fig:adaptation}
%https://app.diagrams.net/#G135w5Bk8OSB7Ak4hxEaaF2DUmZEoIWWbY
\vspace{-15pt}
\end{figure}

There are broadly two categories of methods in adapting LLMs for SR~\cite{yuan2023go}.
The first category of methods, referred to as top tuning (\TT), adapts LLMs by learning an adapter on top of the LLM as illustrated in Figure~\ref{fig:tlt}.
\TT allows efficient adaptation as it does not require both forward and backward propagation across LLMs in training 
%\hl{if we pre-calculate the outputs of LLMs (i.e., $\mathbf{h}^n_i$) for each item text.}
if the outputs of LLMs (i.e., $\mathbf{h}^n_i$) for each item text are pre-calculated.
However, {\TT} could suffer from limited effectiveness (i.e., recommendation performance) as it is essentially equivalent to adapting only the top layer of LLMs. 
%while fixing all the other layers.
%\xia{this is not deep to the point -- why only top layer is not sufficient?}
As suggested in the literature~\cite{shim2021understanding}, different Transformer layers~\cite{vaswani2017attention} in LLMs could capture diverse semantics in item texts.
Thus, adapting only the top layer could lose valuable information in other layers, leading to sub-optimal recommendation performance.
Conversely, the second category of methods, denoted as multi-layer tuning (\FT), enables effective adaptation by adapting potentially all the layers in LLMs via fine-tuning or parameter-efficient fine-tuning (PEFT)~\cite{he2021towards}.
Nevertheless, both fine-tuning and PEFT methods are notoriously  computation-intensive~\cite{liao2023make}.
Consequently, \FT significantly underperforms \TT in terms of efficiency~\cite{yuan2023go}.
Figure~\ref{fig:mt} illustrates \FT in which LLMs are adapted using PEFT.

With the prosperity of LLM-enhanced SR methods, developing both efficient and effective LLM adaptation methods for SR has emerged as a critical research challenge.
To tackle this challenge, in this paper, we introduce a novel side sequential network adaptation method, denoted as \method, for SR.
\method features three key designs.
First, \method learns adapters separate from LLMs, and keeps all the pre-trained parameters within LLMs fixed.
As a result, similar to \TT, with pre-calculated outputs from LLMs (e.g., $\mathbf{h}_i^n$), 
\method could
avoid forward and backward propagation across LLMs during training, and thus, enable efficient adaptation. 
%by pre-calculating outputs from LLMs.
%
Second, distinct from \TT that adapts only the top layer of LLMs, \method adapts the top-$a$ layers jointly to allow effective adaptation.
Third, \method integrates adapters sequentially via a gated recurrent unit (\GRU) network for better effectiveness. 
%\xia{you should mention GRUs right here
%to point out the difference from \FT}.
%
As illustrated in Figure~\ref{fig:mt}, \FT adds adapters within each layer of LLMs.
This design inherently integrates adapters in a sequential manner.
As will be shown in our analysis (Table~\ref{tbl:analysis}), 
the sequential integration of adapters significantly contributes to the effectiveness.
Thus, \method learns an additional \GRU network to sequentially integrate the adapters, thereby enhancing recommendation performance.

We compare \method against five state-of-the-art baseline methods on five benchmark datasets using three LLMs. 
The experimental results demonstrate that \method allows both efficient and effective LLM adaptation for SR.
Particularly, \method outperforms all the baseline methods across the three LLMs with a remarkable average improvement of as much as 8.5\% compared to the best-performing baseline method over the benchmark datasets.
\method also achieves remarkable improvement in terms of run-time efficiency (e.g., 30.8x speedup) and memory efficiency (e.g., 97.7\% reduction in GPU memory usage) compared to the best-performing \FT methods. 
Our analysis and parameter study demonstrates the effectiveness of integrating adapters sequentially, and the effectiveness of adapting top-$a$ layers of LLMs together, respectively.
%\xia{need numbers from the results here. }
%
For better reproducibility, we release the processed datasets and the source code in Google Drive~\footnote{\url{https://drive.google.com/drive/folders/1dKUKLm8s_inzWQORYJFNmeTquCEHyXWi?usp=sharing}}.

%%%%%%%%%%%%%%%%%%%%%%%%%%%%%%%%%%%%%%%%%%%%%%%%%
\section{Related Work}
\label{sec:literature}
%%%%%%%%%%%%%%%%%%%%%%%%%%%%%%%%%%%%%%%%%%%%%%%%%

%**************************************************
\subsection{Sequential Recommendation}
\label{sec:literature:sr}
%**************************************************

Numerous SR methods have been developed in the last few years, particularly, leveraging neural networks and attention mechanisms.
For example, Hidasi~\etal~\cite{hidasi2015session} utilizes \GRUs to capture users' latent intent from their historical interactions. 
Tang~\etal~\cite{tang2018personalized} leverages a convolutional neural network (\CNN) to capture the union-level patterns among items for better user intent modeling.
Besides neural networks, attention mechanisms have also been widely employed for SR.
Ma~\etal~\cite{ma2019hierarchical} learns attention weights to differentiate the importance of items, enhancing the estimate of users' intent.
Kang~\etal~\cite{kang2018self} develops a self-attention-based SR method \SASRec, in which self-attention mechanisms are utilized to capture the interactions between individual items for better user intent modeling.
%
%\SASRec also stacks multiple self-attention layers to capture users' intent in a recursive manner.
%

Recently, increasing efforts have been dedicated to incorporating LLMs for SR.
For example, Hou~\etal~\cite{hou2022towards} adapts the top layer of LLMs for SR using the mixture-of-expert mechanism (\TTMoE), and learns transferable recommendation models based on the adapted item embeddings.
Yuan~\etal~\cite{yuan2023go} shows that item embeddings generated from LLMs could be more expressive than those learned solely based on user interactions, and thus, lead to better recommendation performance.
Yuan~\etal~\cite{yuan2023go} also shows that adapting the top layer of LLMs using multilayer perceptrons (\TTMLP) underperforms \FT methods by a considerable margin.

%**************************************************
\subsection{Parameter-efficient Fine-tuning (PEFT)}
\label{sec:literature:peft}
%**************************************************

As LLMs have demonstrated impressive performance in context understanding, numerous PEFT methods have been developed to adapt LLMs for specific tasks in a parameter-efficient manner.
Particularly, Hu~\etal~\cite{hu2021lora} develops a low-rank adaptation method \LoRA, 
which learns rank decomposition matrices in each LLM layer, while fixing all the pre-trained parameters within LLMs for PEFT. 
Zaken~\etal~\cite{zaken2021bitfit} develops \BitFit, in which only the bias-terms of LLMs are adapted for the task of interest.
Zhang~\etal~\cite{zhang2023adaptive} develops \ADALoRA, which captures the importance of the pre-trained parameters within LLMs, and prioritizes the adaptation of parameters based on their importance.

%%%%%%%%%%%%%%%%%%%%%%%%%%%%%%%%%%%%%%%%%%%%%%%%%
\section{Definition and Notations}
\label{sec:notation}
%%%%%%%%%%%%%%%%%%%%%%%%%%%%%%%%%%%%%%%%%%%%%%%%%

\begin{figure}[t]
	%\vspace{-25pt}
	\includegraphics[width=\linewidth]{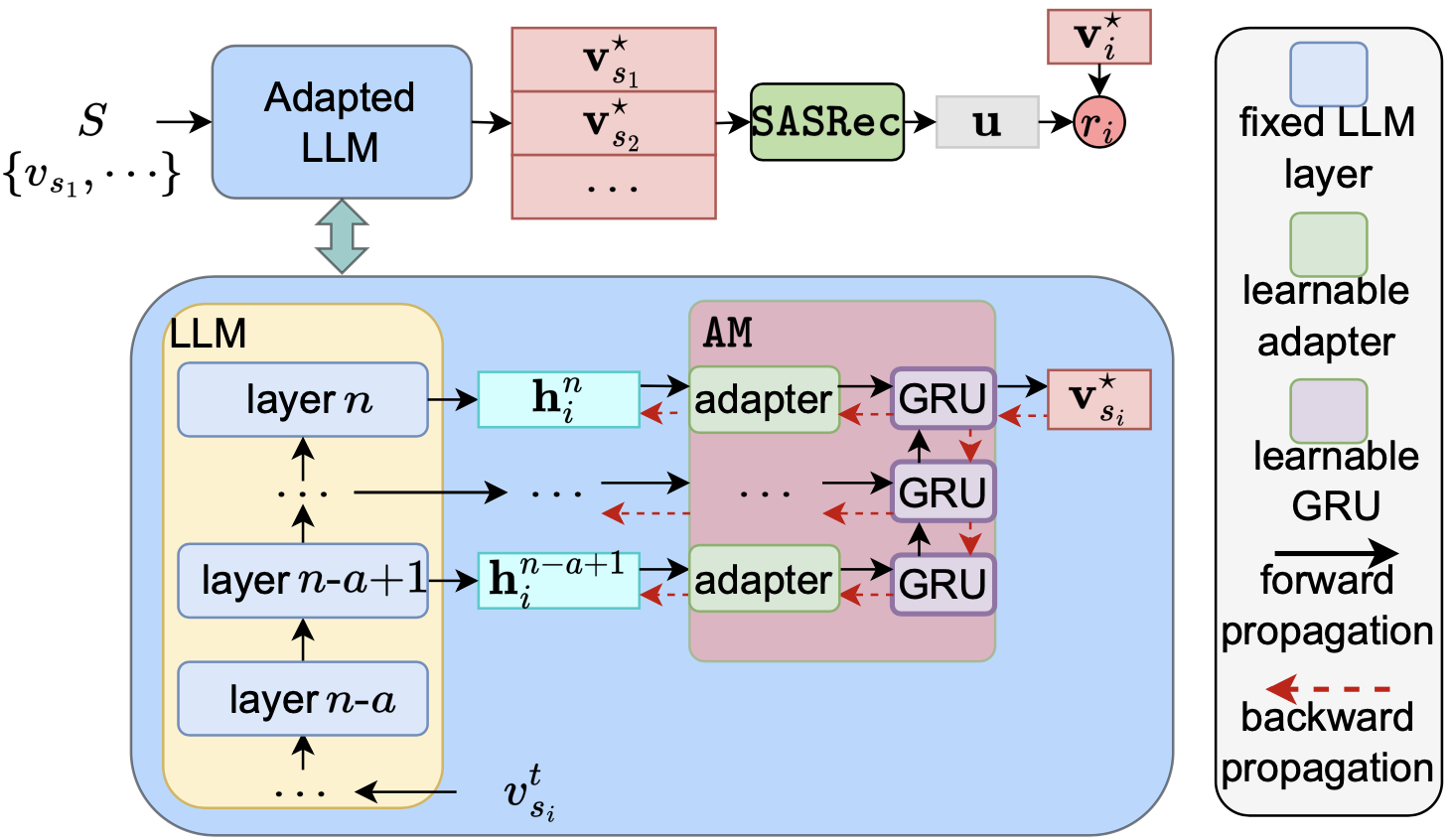}
    %https://app.diagrams.net/#G1Rrk4ITghENYdGuEuEYTAt6dY9Ik7HBl8
	\vspace{-10pt}
	\caption{Overall Architecture of \method. \method jointly adapts the top-$a$ layers of LLMs, and integrates adapters sequentially using a GRU network for effective adaptation.
    In addition, \method could avoid forward and backward propagation across LLMs to enable efficient adaptation if the outputs of LLMs (e.g., $\mathbf{h}_i^n$, $\mathbf{h}_i^{n-1}$) are pre-calculated.}
    %\xia{increase font size; some colors are too bright; add layer $n-a-1$ without going 
	%through AM...}}
	\label{fig:architecture}
	\vspace{-15pt}
\end{figure}

In this paper, we denote the set of all the users as $\mathbb{U} = \{u_1, u_2, \cdots\}$, where $u_j$ is the $j$-th user and $|\mathbb{U}|$ is the total number of users.
We also denote the set of all the items as $\mathbb{V} = \{v_1, v_2, \cdots\}$, 
where $v_i$ is the $i$-th item
and $|\mathbb{V}|$ is the total number of items.
We denote the text (e.g., title) of $v_i$ as $v_i^t$.
In this paper, we represent the historical interactions of $u_j$ as a sequence $S_j = \{v_{s_1}(j), v_{s_2}(j), \cdots\}$ in which $v_{s_t}(j)$ is the $t$-th interacted item in $S_j$.
Given $S_j$, 
we denote the ground-truth next item with which $u_j$ will interact as $v_{g}(j)$.
%
%The goal of \method is to adapt LLMs to correctly recommend $v_{g}(j)$ for $u_j$.
%
When no ambiguity arises, we will eliminate $j$ in $S_j$, $v_{s_t}(j)$ and $v_{g}(j)$.
We use uppercase letters to denote matrices, lower-case bold letters to
denote row vectors and lower-case non-bold letters to represent scalars.

%%%%%%%%%%%%%%%%%%%%%%%%%%%%%%%%%%%%%%%%%%%%%%%%%
\section{Method}
\label{sec:method}
%%%%%%%%%%%%%%%%%%%%%%%%%%%%%%%%%%%%%%%%%%%%%%%%%

Figure~\ref{fig:architecture} presents the overall architecture of \method.
As shown in Figure~\ref{fig:architecture},
\method adapts LLMs by learning adapters for each of the top-$a$ layers of LLMs.
Different from that in \FT, the adapters in \method are separate from LLMs to allow efficient adaptation.
\method also learns a \GRU network to sequentially integrate the adapters for enhanced recommendation performance.
%
%\method learns adapters separate from LLMs, and endows sequential dependencies into the adapters to enable both efficient and effective adaptation.
%
In what follows, we present \method in detail.
%\xia{present Figure 1 and Figure 2 side by side so it is easier to compare. }

%**************************************************
\subsection{LLM Adaptation}
\label{sec:method:llm}
%**************************************************

\method learns an adaptation module (\AM) to adapt LLMs for SR, while fixing all the pre-trained parameters within LLMs.
Particularly, \AM comprises $a$ adapters to adapt the top-$a$ layers of LLMs.
\AM also includes a \GRU network to integrate these adapters sequentially.

%++++++++++++++++++++++++++++++++++++++++++++++++++
\subsubsection{Adapter Design}
\label{sec:method:adapter}
%++++++++++++++++++++++++++++++++++++++++++++++++++

Existing work~\cite{yuan2023go,wang2022transrec} typically implements adapters using multilayer perceptrons (MLPs).
However, as suggested in a recent work~\cite{chen2022towards}, mixture-of-export mechanisms (MoEs)~\cite{shazeer2017outrageously} could better capture the clustering structures within items compared to MLPs, enabling more expressive item embeddings.
Thus, \method implements adapters using MoEs.
Specifically, 
each adapter in \method comprises a projection layer and a routing layer.
Within each projection layer, \method has $n_p$ projection heads.
\method implements each projection head using parametric whitening~\cite{huang2021whiteningbert} as follows:
\begin{equation}
    \label{eqn:proj}
    \mathbf{e}^{m}_i(k) = \left( \mathbf{h}_i^{n-m} - \mathbf{b}^m_k \right ) W^m_k,
\end{equation}
where $\mathbf{h}_i^{n-m}$ is the embedding of item text $v_i^t$ generated from the $(n\text{-}m)$-th layer of 
the LLM ($m \in \{0, 1, \cdots, a\text{-}1\}$), and 
$\mathbf{e}^{m}_i(k)$ is the projected embedding from the $k$-th projection head of the $m$-th adapter.
$\mathbf{b}^m_k$ is a learnable 
shift parameter, 
and $W_k^m$ is a learnable projection parameter.
%
%Following BERT~\cite{devlin2018bert}, we prepend a special ``CLS" token to each item text, and utilize the embedding of this token generated from LLMs to represent the item text.
%

\method utilizes a Gaussian routing to aggregate the projected embedding from each projection head as follows:
\begin{equation}
    \label{eqn:aggregate}
    \mathbf{z}_i^m = \text{Softmax} (\boldsymbol{\alpha}_i^m) [\mathbf{e}_i^m(1); \mathbf{e}_i^m(2); \cdots; \mathbf{e}_i^m(n_p)],
\end{equation}
where $\mathbf{z}_i^m$ is the output from the $m$-th adapter, and \mbox{$\boldsymbol{\alpha}_i^m \sim \mathcal{N}(\boldsymbol{\mu}_i^m, \text{diag}((\boldsymbol{\sigma}_i^m)^2))$} 
is the weights learned to aggregate the projected embeddings.
\method parameterizes the mean $\boldsymbol{\mu}_i^m$ and standard deviation $\boldsymbol{\sigma}_i^m$ of the Gaussian distribution as follows:
\begin{equation}
    \label{eqn:mean}
    \boldsymbol{\mu}^m_i = \mathbf{h}_i^{n-m} B^m 
    \; \text{and} \;
    \boldsymbol{\sigma}^m = \text{Softplus} (\mathbf{h}_i^{n-m} U^m), 
\end{equation}
where 
Softplus is the activation function to 
enable a positive standard deviation;
$B^m$ and 
$U^m$ are learnable parameters in the $m$-th adapter. %\xia{how about $U$ and $V$ instead of $B$ and $U$ in the equations?}.

%++++++++++++++++++++++++++++++++++++++++++++++++++
\subsubsection{Sequential Integration of Adapters}
\label{sec:method:rnn}
%++++++++++++++++++++++++++++++++++++++++++++++++++

%Figure~\ref{fig:mt} illustrates the sequential dependencies exhibit in the adapters learned in \FT.
%
%These dependencies could significantly contribute to the effectiveness of adaptation.
%
%To endow sequential dependencies into adapters as in \FT, \method learns a \GRU network to integrate the adapters sequentially, and generate adapted item embeddings as follows
%
As shown in Figure~\ref{fig:architecture}, \method learns a \GRU network to sequentially integrate the adapters as follows:
\begin{equation}
    \label{eqn:gru}
    \mathbf{v}_i^\star = \GRU( \GRU(\GRU(\mathbf{0}, \mathbf{z}_i^0), \mathbf{z}_i^1) \cdots, \mathbf{z}_i^{a-1}),
\end{equation}
%\xia{why $\mathbf{z}_i^k$, not $\mathbf{z}_i^{n-a}$? also, double check the equation -- it looks inaccurate...}
where $\mathbf{v}_i^\star$ is the adapted embedding for item $v_i$.
We initiate the memory cell in \GRU with a zero vector $\mathbf{0}$ following the literature~\cite{hidasi2015session}.
As illustrated in Figure~\ref{fig:mt}, \FT stacks adapters inside LLMs, and thus, inherently integrates adapters in a sequential manner.
Our own analysis (Table~\ref{tbl:analysis}) demonstrates that this property could significantly contribute to the recommendation performance.
%
%\xia{what is the difference between your sequential integration and {\FT}'s?}
%
Thus, besides adapters, \method learns an additional GRU network to sequentially integrate the adapters for enhanced recommendation performance.

%**************************************************
\subsection{User Intent Modeling and Recommendation Generation}
\label{sec:method:intent}
%**************************************************

We utilize the widely used \SASRec model~\cite{kang2018self} in \method to capture users' intent and generate recommendations.
%
%However, it should be noted that \method serves as a general framework that is also compatible with other user intent modeling methods.
%
%\SASRec is a self-attention-based~\cite{vaswani2017attention} 
%method, which learns temporal patterns using position embeddings, and 
%stacks multiple self-attention layers 
%to aggregate interacted items and 
%estimate users' preference in an iterative manner.
%
Specifically, given the interaction 
sequence of user $u_j$: \mbox{$S_j = \{v_{s_1}(j), v_{s_2}(j), \cdots\}$}, 
we generate adapted embeddings for each of the interacted items, and stack these embeddings to generate an embedding matrix for $S_j$ as follows:
\begin{equation}
    \label{eqn:matrix}
    M_j = [\mathbf{v}_{s_1}^\star(j); \mathbf{v}_{s_2}^\star(j); \cdots], 
\end{equation}
where $\mathbf{v}_{s_t}^\star(j)$ is the adapted embedding of 
item $v_{s_t}(j)$.
Given $M_j$, we model users' intent using \SASRec:
\begin{equation}
    \label{eqn:intent}
    \mathbf{u}_j = \SASRec(M_j),
\end{equation}
where $\mathbf{u}_j$ represents the intent of user $u_j$.
With $\mathbf{u}_j$, the recommendation score of item $v_i$ is calculated as follows:
\begin{equation}
    \label{eqn:score}
    r_{ij} = \cos(\mathbf{u}_j, {\mathbf{v}_i^*}),
\end{equation}
where $\cos(\cdot)$ is the cosine similarity and $r_{ij}$ is the recommendation score of $u_j$ on $v_i$.
\method recommends the top-$k$ items of the highest recommendation scores.

%**************************************************
\subsection{Network Training}
\label{sec:method:training}
%**************************************************
 
We utilize the cross-entropy objective function
to minimize the negative log-likelihood of correctly recommending the ground truth next item as follows:
\begin{equation}
    \begin{aligned}
    \label{eqn:obj}
    \min \limits_{\boldsymbol{\Theta}} \sum_{u_j \in \mathbb{U}} 
    \frac{\exp(\cos(\mathbf{u}_j, \mathbf{v}_{g}^*(j)) / \tau)}
        {\sum_{v_{i} \in \mathbb{N}_j} 
        \exp(\cos(\mathbf{u}_j, \mathbf{v}_{i}^*) / \tau)},
    \end{aligned}
\end{equation}
%\xia{the loss is not correct! fix it}
where $\mathbb{U}$ is the set of all the users; 
$\boldsymbol{\Theta}$ is the set of learnable parameters in \method (i.e., parameters in \AM and \SASRec); 
$\exp(\cdot)$ is the exponential function;
$\mathbf{v}_{g}^*(j)$ is the adapted embedding of the ground-truth next item of $u_j$;
$\mathbb{N}_j$ is the set of randomly sampled negative items of $u_j$;
$\tau$ is a hyper-parameter to scale the cosine similarities~\cite{hou2022towards}.
\method minimizes the objective function using batch optimizations.
For $u_j$ within a particular batch,
\method regards the ground-truth next items from all other users in the same batch as the set of negative items (i.e., $\mathbb{N}_j$) of $u_j$.
All the learnable parameters are randomly initialized, 
and are optimized in an end-to-end manner.
%
%Note that, \method pre-calculates the embedding of each item from the top-$k$ layers of LLMs (i.e., $\mathbf{h}_i^{n-m}$) to enable efficient training.

%%%%%%%%%%%%%%%%%%%%%%%%%%%%%%%%%%%%%%%%%%%%%%%%%
\section{Materials}
\label{sec:materials}
%%%%%%%%%%%%%%%%%%%%%%%%%%%%%%%%%%%%%%%%%%%%%%%%%

%**************************************************
\subsection{Baseline Methods}
\label{sec:materials:baseline}
%**************************************************

We compare \method against five state-of-the-art baseline methods.
Specifically, we compare \method with the state-of-the-art \FT methods \LoRA~\cite{hu2021lora},  \ADALoRA~\cite{zhang2023adaptive}, and full fine-tuning (\Full)~\cite{yuan2023go}.
Besides \FT methods, we also compare \method with \TT methods: \TTMLP~\cite{yuan2023go} 
and \TTMoE~\cite{hou2022towards}, in which MPLs and MoEs are utilized to implement the adapter, respectively.
%
%\TTMLP implements the adapter using MLPs.
%
%In contrast, \TTMoE implements the adapter using MoEs, 
%the same as that in \method.
%
We refer the audience of interest to the Section ``Related Work"
for details of the baseline methods.

\LoRA and \ADALoRA have demonstrated superior performance over a comprehensive set of other methods 
such as \BitFit~\cite{zaken2021bitfit}.
Thus, we compare \method with \LoRA and \ADALoRA instead of the methods that they outperform.
For \LoRA and \ADALoRA, we use the implementation in PEFT~\cite{peft}, a widely used library for parameter-efficient fine-tuning.
% \xia{fine tuning}.
%
We implement \Full, \TTMLP and \TTMoE using Pytorch-lighting and Transformers~\cite{wolf-etal-2020-transformers}.
Transformers is a prevalent library for fine tuning LLMs.
For a fair comparison, we employ \SASRec, the user intent modeling method utilized in \method, for all the baseline methods.

%**************************************************
\subsection{Datasets}
\label{sec:materials:datasets}
%**************************************************

\begin{table}
\footnotesize
  \caption{Dataset Statistics}
  \centering
  \vspace{-10pt}
  \label{tbl:dataset}
  \begin{threeparttable}
      \begin{tabular}{
    @{\hspace{6pt}}l@{\hspace{6pt}}
	@{\hspace{6pt}}r@{\hspace{6pt}}          
	@{\hspace{6pt}}r@{\hspace{6pt}}
	@{\hspace{6pt}}r@{\hspace{6pt}}
	@{\hspace{6pt}}r@{\hspace{6pt}}
    @{\hspace{6pt}}r@{\hspace{6pt}}
	}
        \toprule
        dataset & \#users & \#items & \#intrns & {\#intrns/u} & {\#intrns/i}\\
        \midrule
        \SCI    &  8,442 & 4,385 &  59,427 & 7.0 & 13.6\\
        \Pantry & 13,101 & 4,898 & 126,962 & 9.7 & 25.9\\
        \Tools  &  8,959 & 5,871 & 145,681 &16.3 & 24.8\\
        \Toys   & 11,803 & 8,569 & 206,103 &17.5 & 24.1\\
        \INS    & 24,962 & 9,964 & 208,926 & 8.4 & 21.0\\
        \bottomrule
      \end{tabular}
      \begin{tablenotes}[normal,flushleft]
      \begin{footnotesize}
      \item 
      In this table, 
      the column ``\#users", ``\#items" and ``\#intrns" shows the number of 
      users, items and user-item interactions, respectively. 
      The column ``\#intrns/u" has the average number of interactions for each user. 
      The column ``\#intrns/i" has the average number of interactions for each item.  
      \par
      \end{footnotesize}
      \end{tablenotes}
  \end{threeparttable}
  \vspace{-15pt}
\end{table}
%\label{tbl:dataset}

We evaluate \method and baseline methods using five benchmark datasets: 
Amazon-Scientific (\SCI), Amazon-Pantry (\Pantry), Amazon-Tools (\Tools), 
Amazon-Toys (\Toys) and Amazon-Instruments (\INS).
We download all the datasets from Amazon Review~\cite{he2016ups}, which includes users' interactions on different categories of items, and the associated text for each item.
Following the literature~\cite{hou2022towards}, we concatenate title, sub-categories and brand of each item as the item text.
For \SCI, \Pantry and \INS, we keep users and items with at least 5
interactions following the literature~\cite{hou2022towards}.
For \Toys, we keep users and items with at least 10 interactions for efficiency consideration.
For \Tools, we keep users and items with at least 11 interactions to enable fine tuning LLMs using V100 GPUs with 32GB memory.
On all the datasets, for each user, we consider only her most recent 50 items in our experiments following the literature~\cite{kang2018self,hou2022towards}.
Table~\ref{tbl:dataset} presents the statistics of the processed datasets.

%include users’ interactions and
%reviews on different categories of products.

%**************************************************
\subsection{LLMs}
\label{sec:materials:llm}
%**************************************************

We utilize three LLMs in our experiments: \DistillBERT~\cite{sanh2019distilbert}, \DistillRoBERTa~\footnote{\url{https://huggingface.co/distilroberta-base}} and \BERTMedium~\cite{turc2019well}.
\DistillBERT and \DistillRoBERTa are both comprised of 6 Transformer layers and are trained using knowledge distillation with \BERT~\cite{devlin2018bert} and \RoBERTa~\cite{liu2019roberta} as teacher models, respectively.
\BERTMedium is comprised of 8 Transformer layers, and it is trained through a combination of pre-training and knowledge distillation. 
We download all the LLMs from Hugging Face~\cite{wolf-etal-2020-transformers}.
Compared to \BERT and \RoBERTa, \DistillBERT, \DistillRoBERTa and \BERTMedium require significantly fewer resources in fine-tuning, while achieving competitive performance.
Thus, due to the resource limitation, we utilize \DistillBERT, \DistillRoBERTa and \BERTMedium instead of \BERT and \RoBERTa in our experiments.
In line with the literature~\cite{hou2022towards,devlin2018bert}, for all the LLMs, we incorporate a special ``CLS" token as a prefix to each item text $v_i^t$, and use the embedding of this token as the embedding of $v_i^t$ in each layer of LLMs.

%**************************************************
\subsection{Experimental Protocol}
\label{sec:materials:protocol}
%**************************************************

%+++++++++++++++++++++++++++++++++++++++++++++++++++++++++++
\subsubsection{Training, Validation and Testing Set}
\label{sec:materials:protocol:training}
%+++++++++++++++++++++++++++++++++++++++++++++++++++++++++++

For each user, following the literature~\cite{fan2022sequential}, we use her last and second last interaction for testing and validation, respectively.
The other interactions are used for training.
Following \SASRec~\cite{kang2018self}, we split each training sequence to sub-sequences and train \method and all the baseline methods using all the sub-sequences.
%

%+++++++++++++++++++++++++++++++++++++++++++++++++++++++++++
\subsubsection{Evaluation Metrics}
\label{sec:materials:protocol:evaluation}
%+++++++++++++++++++++++++++++++++++++++++++++++++++++++++++
 
We evaluate all the methods using the widely used evaluation metrics Recall@$k$ (\RK) and NDCG@$k$ (\NK).
We refer the audience of interest to Peng~\etal~\cite{peng2021ham} for the detailed definations.
%the detailed definitions of \RK and \NK.

%**************************************************
\subsection{Hyper-parameter Tuning}
\label{sec:materials:parameter}
%**************************************************

We tune hyper-parameters for \method and all the baseline methods using
grid search.
For all the methods, we use the best-performing hyper-parameters in terms of \RT on the validation set for testing.
We use Adam optimizer with learning rate
3e-4 for all the methods.
For \LoRA, \ADALoRA and \Full, we set the batch size $b$ as 32 and accumulate gradients for every 8 batches~\cite{ag}.
We observe that larger batch sizes could lead to the out-of-memory issue on GPUs.
For \TTMLP, \TTMoE and \method, we search $b$ in $\{32, 64, 128, 256\}$.
For \LoRA and \ADALoRA, we search the dropout probability $p$ in $\{0.1, 0.2, 0.3\}$.
For \Full, we search the number of top layers to fully fine-tune, denoted as $n_t$, in $\{2, 4, 6\}$.
For \TTMLP, we search the number of MLP layers $n_l$ in $\{1, 2, 3\}$.
For \method, we search the number of top layers to adapt ($a$) in $\{2, 3\}$.
We set the number of projection heads in each adapter ($n_p$) as 8.
We also set the scaling parameter $\tau$ (Equation~\ref{eqn:obj}) as 0.07.
Following the literature~\cite{hou2022towards}, 
we set both the number of self-attention layers and the
number of attention heads in \SASRec as 2. 
We also set the dimension for item embeddings as 256 for all the methods.
We report the best-performing hyper-parameters of \method and all the baseline methods in Table~\ref{tbl:para_distillbert}, Table~\ref{tbl:para_distillroberta} and Table~\ref{tbl:para_bertmedium} (Appendix).

To accelerate training, we train \LoRA, \ADALoRA and \Full using 2 GPUs.
%
%This strategy implicitly doubles the batch size in these methods.
%
We use the 16-bit precision for all the methods during training to lower the memory consumption on GPUs.
We train \LoRA, \ADALoRA and \Full using V100 GPUs with 32GB memory on \SCI, \Pantry and \Tools.
On \Toys and \INS, we train \LoRA, \ADALoRA and \Full using A100 GPUs with 80GB memory.
We train \TTMLP, \TTMoE and \method on all the datasets using V100 GPUs with 32GB memory.
As we have limited access to A100 GPUs, we
cannot conduct all the experiments using A100 GPUs.
%

%%%%%%%%%%%%%%%%%%%%%%%%%%%%%%%%%%%%%%%%%%%%%%%%%
\section{Experimental Results}
\label{sec:results}
%%%%%%%%%%%%%%%%%%%%%%%%%%%%%%%%%%%%%%%%%%%%%%%%%

%**************************************************
\subsection{Recommendation Performance}
\label{sec:results:recommendation}
%**************************************************

\begin{table}
\footnotesize
  \caption{Overall Performance (\DistillBERT)}
  \centering
  \vspace{-10pt}
  \label{tbl:distillbert}
  \begin{threeparttable}
      \begin{tabular}{
        @{\hspace{0pt}}l@{\hspace{3pt}}
	  @{\hspace{3pt}}l@{\hspace{3pt}}
	  @{\hspace{3pt}}r@{\hspace{3pt}}
	  @{\hspace{3pt}}r@{\hspace{3pt}}
	  @{\hspace{3pt}}r@{\hspace{3pt}}
	  @{\hspace{3pt}}r@{\hspace{3pt}}
        @{\hspace{3pt}}r@{\hspace{3pt}}
        @{\hspace{3pt}}r@{\hspace{0pt}}
      }
      \toprule
      Dataset & Metric & \LoRA & \ADALoRA & \Full & \TTMLP & \TTMoE & \method\\
      \midrule
      \multirow{4}{*}{\SCI}
      & \RT & \textbf{0.1070} & 0.1025 & 0.0974 & 0.0964 & 0.0978 & \underline{0.1059}\\
      & \RF & 0.1931 & \underline{0.1950} & 0.1821 & 0.1863 & 0.1857 & \textbf{0.1986}\\
      \cline{2-8}
      & \NT & 0.0670 & \underline{0.0678} & 0.0623 & 0.0621 & 0.0630 & \textbf{0.0692}\\
      & \NF & 0.0856 & \underline{0.0877} & 0.0808 & 0.0817 & 0.0823 & \textbf{0.0892}\\
      \midrule
      \multirow{4}{*}{\Pantry}
      & \RT & \textbf{0.0588} & \underline{0.0585} & 0.0380 & 0.0521 & 0.0554 & 0.0581\\
      & \RF & \underline{0.1230} & \textbf{0.1239} & 0.0982 & 0.1142 & 0.1149 & 0.1220\\
      \cline{2-8}
      & \NT & \underline{0.0325} & \textbf{0.0326} & 0.0192 & 0.0283 & 0.0305 & 0.0323\\
      & \NF & \underline{0.0464} & \textbf{0.0468} & 0.0321 & 0.0418 & 0.0433 & 0.0460\\
      \midrule
      \multirow{4}{*}{\Tools}
      & \RT & \underline{0.0670} & 0.0646 & 0.0622 & 0.0646 & 0.0623 & \textbf{0.0673}\\
      & \RF & \underline{0.1316} & 0.1299 & 0.1269 & 0.1303 & 0.1235 & \textbf{0.1352}\\
      \cline{2-8}
      & \NT & \underline{0.0450} & 0.0432 & 0.0432 & 0.0441 & 0.0428 & \textbf{0.0460}\\
      & \NF & \underline{0.0589} & 0.0571 & 0.0571 & 0.0581 & 0.0560 & \textbf{0.0605}\\
      \midrule
      \multirow{4}{*}{\Toys}
      & \RT & \underline{0.0802} & 0.0737 & 0.0750 & 0.0800 & 0.0779 & \textbf{0.0855}\\
      & \RF & \underline{0.1884} & 0.1815 & 0.1694 & 0.1828 & 0.1744 & \textbf{0.1910}\\
      \cline{2-8}
      & \NT & \underline{0.0452} & 0.0413 & 0.0421 & 0.0447 & 0.0435 & \textbf{0.0499}\\
      & \NF & \underline{0.0687} & 0.0645 & 0.0625 & 0.0669 & 0.0644 & \textbf{0.0727}\\
      \midrule
      \multirow{4}{*}{\INS}
      & \RT & \underline{0.0808} & 0.0800 & 0.0723 & 0.0780 & 0.0797 & \textbf{0.0841}\\
      & \RF & \underline{0.1520} & \textbf{0.1523} & 0.1367 & 0.1416 & 0.1403 & 0.1516\\
      \cline{2-8}
      & \NT & 0.0583 & 0.0584 & 0.0546 & 0.0588 & \underline{0.0610} & \textbf{0.0641}\\
      & \NF & 0.0737 & 0.0739 & 0.0683 & 0.0726 & \underline{0.0741} & \textbf{0.0786}\\
      \bottomrule
      \end{tabular}
      \begin{tablenotes}[normal,flushleft]
      \begin{footnotesize}
      \item
      In this table, the best performance and the second-best performance on each dataset is in \textbf{bold} and \underline{underlined}, respectively.
      \par
      \end{footnotesize}
      %\end{scriptsize}
      \end{tablenotes}
  \vspace{-10pt}
  \end{threeparttable}
\end{table}
%\label{tbl:distillbert}

\begin{table}
\footnotesize
  \caption{Overall Performance (\DistillRoBERTa)}
  \centering
  \vspace{-10pt}
  \label{tbl:distillroberta}
  \begin{threeparttable}
      \begin{tabular}{
        @{\hspace{0pt}}l@{\hspace{3pt}}
	  @{\hspace{3pt}}l@{\hspace{3pt}}
	  @{\hspace{3pt}}r@{\hspace{3pt}}
	  @{\hspace{3pt}}r@{\hspace{3pt}}
	  @{\hspace{3pt}}r@{\hspace{3pt}}
	  @{\hspace{3pt}}r@{\hspace{3pt}}
        @{\hspace{3pt}}r@{\hspace{3pt}}
        @{\hspace{3pt}}r@{\hspace{3pt}}
        @{\hspace{3pt}}r@{\hspace{3pt}}
      }
      \toprule
      Dataset & Metric & \LoRA & \ADALoRA & \Full & \TTMLP & \TTMoE & \method\\
      \midrule
      \multirow{4}{*}{\SCI}
      & \RT & \textbf{0.1104} & \underline{0.1039} & 0.1013 & 0.1018 & 0.0962 & 0.1014\\
      & \RF & \textbf{0.1941} & 0.1905 & 0.1831 & 0.1864 & 0.1829 & \underline{0.1911}\\
      \cline{2-8}
      & \NT & \textbf{0.0681} & \underline{0.0665} & 0.0649 & 0.0627 & 0.0613 & 0.0635\\
      & \NF & \textbf{0.0862} & \underline{0.0854} & 0.0827 & 0.0811 & 0.0802 & 0.0829\\
      \midrule
      \multirow{4}{*}{\Pantry}
      & \RT & \underline{0.0553} & \textbf{0.0569} & 0.0374 & 0.0496 & 0.0517 & 0.0540\\
      & \RF & \textbf{0.1242} & \underline{0.1230} & 0.0954 & 0.1022 & 0.1087 & 0.1190\\
      \cline{2-8}
      & \NT & \underline{0.0299} & \textbf{0.0316} & 0.0193 & 0.0269 & 0.0290 & 0.0291\\
      & \NF & \underline{0.0447} & \textbf{0.0458} & 0.0318 & 0.0383 & 0.0414 & 0.0432\\
      \midrule
      \multirow{4}{*}{\Tools}
      & \RT & \underline{0.0695} & 0.0645 & 0.0629 & 0.0616 & 0.0646 & \textbf{0.0722}\\
      & \RF & \underline{0.1340} & 0.1327 & 0.1299 & 0.1298 & 0.1231 & \textbf{0.1349}\\
      \cline{2-8}
      & \NT & \underline{0.0477} & 0.0436 & 0.0441 & 0.0424 & 0.0449 & \textbf{0.0503}\\
      & \NF & \underline{0.0616} & 0.0582 & 0.0584 & 0.0570 & 0.0574 &  \textbf{0.0637}\\
      \midrule
      \multirow{4}{*}{\Toys}
      & \RT & 0.0731 & 0.0685 & 0.0773 & 0.0702 & \underline{0.0793} & \textbf{0.0859}\\
      & \RF & \underline{0.1811} & 0.1705 & 0.1710 & 0.1715 & 0.1722 & \textbf{0.1847}\\
      \cline{2-8}
      & \NT & 0.0415 & 0.0382 & 0.0440 & 0.0391 & \underline{0.0448} & \textbf{0.0491}\\
      & \NF & \underline{0.0648} & 0.0604 & 0.0642 & 0.0610 & \underline{0.0648} & \textbf{0.0705}\\
      \midrule
      \multirow{4}{*}{\INS}
      & \RT & 0.0798 & 0.0788 & 0.0761 & 0.0798 & \underline{0.0851} & \textbf{0.0875}\\
      & \RF & 0.1509 & \underline{0.1515} & 0.1414 & 0.1463 & 0.1463 & \textbf{0.1560}\\
      \cline{2-8}
      & \NT & 0.0560 & 0.0563 & 0.0568 & 0.0619 & \underline{0.0650} & \textbf{0.0675}\\
      & \NF & 0.0712 & 0.0719 & 0.0708 & 0.0761 & \underline{0.0781} & \textbf{0.0823}\\
      \bottomrule
      \end{tabular}
      \begin{tablenotes}[normal,flushleft]
      \begin{footnotesize}
      \item
      In this table, the best performance and the second-best performance on each dataset is in \textbf{bold} and \underline{underlined}, respectively.
      \par
      \end{footnotesize}
      %\end{scriptsize}
      \end{tablenotes}
  \vspace{-5pt}
  \end{threeparttable}
\end{table}

%\label{tbl:distillroberta}

\begin{table}[!t]
\footnotesize
  \caption{Overall Performance (\BERTMedium)}
  \centering
  \vspace{-10pt}
  \label{tbl:bertmedium}
  \begin{threeparttable}
      \begin{tabular}{
        @{\hspace{0pt}}l@{\hspace{3pt}}
	  @{\hspace{3pt}}l@{\hspace{3pt}}
	  @{\hspace{3pt}}r@{\hspace{3pt}}
	  @{\hspace{3pt}}r@{\hspace{3pt}}
	  @{\hspace{3pt}}r@{\hspace{3pt}}
	  @{\hspace{3pt}}r@{\hspace{3pt}}
        @{\hspace{3pt}}r@{\hspace{3pt}}
        @{\hspace{3pt}}r@{\hspace{3pt}}
      }
      \toprule
      Dataset & Metric & \LoRA & \ADALoRA & \Full & \TTMLP & \TTMoE & \method\\
      \midrule
      \multirow{4}{*}{\SCI}
      & \RT & 0.1034 & 0.0989 & 0.0968 & \underline{0.1042} & 0.0980 & \textbf{0.1077}\\
      & \RF & \underline{0.1925} & \textbf{0.1931} & 0.1855 & 0.1889 & 0.1857 & \underline{0.1925}\\
      \cline{2-8}
      & \NT & 0.0652 & 0.0620 & 0.0635 & \underline{0.0654} & 0.0641 & \textbf{0.0692}\\
      & \NF & \underline{0.0846} & 0.0824 & 0.0828 & 0.0838 & 0.0833 & \textbf{0.0876}\\
      \midrule
      \multirow{4}{*}{\Pantry}
      & \RT & 0.0559 & \underline{0.0574} & 0.0439 & 0.0544 & 0.0531 & \textbf{0.0576}\\
      & \RF & \textbf{0.1247} & \underline{0.1243} & 0.1020 & 0.1154 & 0.1096 & 0.1224\\
      \cline{2-8}
      & \NT & 0.0309 & \underline{0.0314} & 0.0236 & 0.0301 & 0.0303 & \textbf{0.0332}\\
      & \NF & 0.0458 & \underline{0.0459} & 0.0361 & 0.0433 & 0.0425 & \textbf{0.0471}\\
      \midrule
      \multirow{4}{*}{\Tools}
      & \RT & 0.0637 & 0.0616 & 0.0623 & 0.0618 & \underline{0.0640} & \textbf{0.0695}\\
      & \RF & \underline{0.1379} & 0.1314 & 0.1271 & 0.1290 & 0.1248 & \textbf{0.1381}\\
      \cline{2-8}
      & \NT & \underline{0.0440} & 0.0415 & 0.0423 & 0.0436 & 0.0438 & \textbf{0.0485}\\
      & \NF & \underline{0.0599} & 0.0564 & 0.0563 & 0.0581 & 0.0569 & \textbf{0.0632}\\
      \midrule
      \multirow{4}{*}{\Toys}
      & \RT & 0.0716 & 0.0651 & 0.0733 & \underline{0.0767} & 0.0737 & \textbf{0.0846}\\
      & \RF & 0.1766 & 0.1716 & 0.1661 & \underline{0.1788} & 0.1697 & \textbf{0.1867}\\
      \cline{2-8}
      & \NT & 0.0394 & 0.0346 & 0.0418 & \underline{0.0425} & 0.0418 & \textbf{0.0489}\\
      & \NF & 0.0620 & 0.0577 & 0.0620 & \underline{0.0646} & 0.0627 & \textbf{0.0711}\\
      \midrule
      \multirow{4}{*}{\INS}
      & \RT & 0.0788 & 0.0785 & 0.0712 & 0.0797 & \underline{0.0809} & \textbf{0.0853}\\
      & \RF & \textbf{0.1532} & \underline{0.1509} & 0.1333 & 0.1453 & 0.1432 & \textbf{0.1532}\\
      \cline{2-8}
      & \NT & 0.0537 & 0.0544 & 0.0527 & 0.0587 & \underline{0.0593} & \textbf{0.0624}\\
      & \NF & 0.0697 & 0.0699 & 0.0659 & \underline{0.0729} & 0.0727 & \textbf{0.0769}\\
      \bottomrule
      \end{tabular}
      \begin{tablenotes}[normal,flushleft]
      \begin{footnotesize}
      \item
      In this table, the best performance and the second-best performance on each dataset is in \textbf{bold} and \underline{underlined}, respectively.
      \par
      \end{footnotesize}
      %\end{scriptsize}
      \end{tablenotes}
  \vspace{-10pt}
  \end{threeparttable}
\end{table}

%\label{tbl:bertmedium}

\begin{table}[!t]
\footnotesize
  \caption{Performance Improvement of \method (\%)}
  \centering
  \vspace{-10pt}
  \label{tbl:improvement}
  \begin{threeparttable}
      \begin{tabular}{
        @{\hspace{0pt}}l@{\hspace{4pt}}
	  @{\hspace{4pt}}l@{\hspace{4pt}}
	  @{\hspace{4pt}}r@{\hspace{4pt}}
	  @{\hspace{4pt}}r@{\hspace{4pt}}
	  @{\hspace{4pt}}r@{\hspace{4pt}}
	  @{\hspace{4pt}}r@{\hspace{4pt}}
        @{\hspace{4pt}}r@{\hspace{0pt}}
      }
      \toprule
      LLMs & Metric & \LoRA & \ADALoRA & \Full & \TTMLP & \TTMoE\\
      \midrule
      \multirow{4}{*}{\DistillBERT}
      & \RT & 1.8\textcolor{white}{$^*$} & 5.6$^*$ & 20.0$^*$ & 8.0$^*$ & 7.3$^*$\\
      & \RF & 1.2\textcolor{white}{$^*$} & 1.8\textcolor{white}{$^*$} & 12.7$^*$ & 5.7$^*$ & 8.0$^*$\\
      \cline{2-7}
      & \NT & 5.0$^*$ & 7.6$^*$ & 24.3$^*$ & 10.1$^*$ & 8.6$^*$\\
      & \NF & 3.7$^*$ & 5.0$^*$ & 18.2$^*$ &  8.1$^*$ & 8.3$^*$\\
      \midrule
      \multirow{4}{*}{\DistillRoBERTa}
      & \RT & 4.1\textcolor{white}{$^*$} & 8.2\textcolor{white}{$^*$} & 17.1$^*$ & 11.5$^*$ & 6.6$^*$\\
      & \RF & 0.1\textcolor{white}{$^*$} & 2.0\textcolor{white}{$^*$} & 10.3$^*$ & 7.4$^*$ & 7.5$^*$\\
      \cline{2-7}
      & \NT & 7.0\textcolor{white}{$^*$} & 10.3\textcolor{white}{$^*$} & 18.6$^*$ & 12.5$^*$ & 5.9$^*$\\
      & \NF & 4.1\textcolor{white}{$^*$} & 6.4\textcolor{white}{$^*$} & 14.2$^*$ & 10.1$^*$ & 6.6$^*$\\
      \midrule
      \multirow{4}{*}{\BERTMedium}
      & \RT & 8.5$^*$ & 12.1$^*$ & 17.8$^*$ & 7.8$^*$ & 9.4$^*$\\
      & \RF & 0.8\textcolor{white}{$^*$} & 2.7\textcolor{white}{$^*$} & 12.0$^*$ & 5.0$^*$ & 8.6$^*$\\
      \cline{2-7}
      & \NT & 12.8$^*$ & 18.0$^*$ & 19.9$^*$ & 9.7$^*$ & 10.1$^*$\\
      & \NF & 7.4$^*$ & 10.8$^*$ & 16.0$^*$ & 7.5$^*$ & 9.2$^*$\\
      \bottomrule
      \end{tabular}
      \begin{tablenotes}[normal,flushleft]
      \begin{footnotesize}
      \item
      In this table, the ${^*}$ indicates that the improvement is statistically significant at 85\% confidence level.
      \par
      \end{footnotesize}
      %\end{scriptsize}
      \end{tablenotes}
  \vspace{-10pt}
  \end{threeparttable}
\end{table}

%\label{tbl:improvement}

Table~\ref{tbl:distillbert}, Table~\ref{tbl:distillroberta} and Table~\ref{tbl:bertmedium} shows the recommendation performance of \method and all the baseline methods on the five benchmark datasets using \DistillBERT, \DistillRoBERTa and \BERTMedium, respectively.
Table~\ref{tbl:improvement} presents the average improvement of \method over each of the baseline method across the five datasets using \DistillBERT, \DistillRoBERTa and \BERTMedium.
In Table~\ref{tbl:improvement}, we test the significance of the improvement using paired t-test.

%\xia{you need to discuss each table seperately first, and then the overall conclusions across tables... it is not clear how you came up with the following conclusions: \method is the best, \LoRA is the second best... you also need to discuss the different LLM models and how and why they have different results on the different datasets...}
Table~\ref{tbl:distillbert}, Table~\ref{tbl:distillroberta} and Table~\ref{tbl:bertmedium} together show that overall, \method is the best-performing method on the five datasets across the three LLMs. 
%\xia{is this accurate? for {\DistillRoBERTa}, it seems it is not the best overall. }
%
For example, as shown in Table~\ref{tbl:distillbert},
using \DistillBERT, 
\method achieves the best performance at \RT and \NT on three and four out of the five datasets, respectively.
Similarly, as shown in Table~\ref{tbl:bertmedium}, using \BERTMedium, \method outperforms all the baseline methods across all the five datasets at both \RT and \NT.
The performance of \method when using \DistillRoBERTa is slightly worse than that of using \DistillBERT and \BERTMedium as shown in Table~\ref{tbl:distillroberta}.
However, as presented in Table~\ref{tbl:improvement}, using \DistillRoBERTa, \method still achieves a considerable average improvement of 4.1\% and 7.0\% at \RT and \NT, respectively, over the best-performing baseline method \LoRA.
%
%In terms of \RT, \method achieves a considerable average improvement of 2.4\%, 2.8\% and 8.5\% using \DistillBERT, \DistillRoBERTa and \BERTMedium, respectively, when compared to the best-performing baseline method \LoRA.
%
%Similarly, at \NT, \method outperforms \LoRA with a significant average improvement of 5.7\% and 12.8\% using \DistillBERT and \BERTMedium, respectively. 
%
These results demonstrate the superior effectiveness of \method in adapting LLMs for SR over the state-of-the-art baseline methods.

%+++++++++++++++++++++++++++++++++++++++++++++++++++++++++++
\subsubsection{Comparison between \method and \LoRA}
\label{sec:results:recommendation:lora}
%+++++++++++++++++++++++++++++++++++++++++++++++++++++++++++

As shown in Table~\ref{tbl:improvement}, overall, \method outperforms the best-performing \FT method \LoRA on the five datasets using the three LLMs.
For example, with \DistillBERT, \method achieves an average improvement of 1.8\%, 1.2\%, 5.0\% and 3.7\% at \RT, \RF, \NT and \NF, respectively, when compared to \LoRA.
\method and \LoRA both integrate adapters in a sequential manner for effective adaptation.
However, while \LoRA stacks adapters over layers of LLMs, 
\method learns adapters separate from LLMs and employs a GRU network for the integration.
Previous work~\cite{he2016deep} shows that gradient propagation across deep networks could be challenging.
Consequently, compared to \LoRA, the shallow architecture in \method could facilitate gradient-based optimization, leading to better performance.
In addition, while \LoRA adapts all the layers in LLMs, \method focuses on adapting only the top-$a$ layers.
As will be shown in Figure~\ref{fig:para}, we empirically find that adapting all the layers of LLMs may degrade the recommendation performance.

%+++++++++++++++++++++++++++++++++++++++++++++++++++++++++++
\subsubsection{Comparing \method to \TTMLP and \TTMoE}
\label{sec:results:recommendation:lora}
%+++++++++++++++++++++++++++++++++++++++++++++++++++++++++++

Table~\ref{tbl:improvement} also shows that compared to the state-of-the-art \TT method \TTMLP and \TTMoE, \method achieves significant average improvement over the five datasets with the three LLMs.
For example, with \DistillBERT, in terms of \RT, \method outperforms both \TTMLP and \TTMoE on all the five datasets, and achieves a significant average improvement of 8.0\% and 7.3\% over \TTMLP and \TTMoE, respectively.
\TTMLP and \TTMoE adapts only the top layer of LLMs, while \method adapts the top-$a$ layers jointly for enhanced recommendation performance.
The superior performance of \method over \TTMLP and \TTMoE demonstrates the importance of adapting top-$a$ layers of LLMs together in enabling effective recommendation.
%\xia{why do you not discuss {\Full}?}

%Table~\ref{tbl:improvement} also shows that overall, \method significantly outperforms the state-of-the-art \TT method \TTMLP and \TTMoE on the five datasets using all the three LLMs.

%**************************************************
\subsection{Comparison on Efficiency}
\label{sec:results:efficiency}
%**************************************************

We further compare the efficiency of \method against the best-performing \FT methods \LoRA and \ADALoRA.
Particularly, in this paper, we focus on the run-time efficiency and memory efficiency of these methods during training.
We assess run-time efficiency based on the run-time per epoch of different methods following the literature~\cite{yuan2023go}, and measure the memory efficiency using the memory usage on GPUs.
We conduct the comparison using the best-performing hyper-parameters of \method, \LoRA and \ADALoRA on different datasets.
To enable a fair comparison, we train all the methods using V100 GPUs with 32GB memory on \SCI, \Pantry and \Tools.
On \Toys and \INS, we train all the methods using A100 GPUs with 80GB memory.
%
%As we have limited access to A100 GPUs, we cannot conduct all the experiments using A100 GPUs.

%+++++++++++++++++++++++++++++++++++++++++++++++++++++++++++
\subsubsection{Comparison on Run-time Efficiency}
\label{sec:results:efficiency:runtime}
%+++++++++++++++++++++++++++++++++++++++++++++++++++++++++++

\begin{table}[!t]
\footnotesize
  \caption{Comparison on Runtime per Epoch (second)}
  \centering
  \vspace{-10pt}
  \label{tbl:runtime}
  \begin{threeparttable}
      \begin{tabular}{
        @{\hspace{0pt}}l@{\hspace{2pt}}
	  @{\hspace{2pt}}l@{\hspace{2pt}}
	  @{\hspace{2pt}}r@{\hspace{2pt}}
	  @{\hspace{2pt}}r@{\hspace{2pt}}
	  @{\hspace{2pt}}r@{\hspace{2pt}}
	  @{\hspace{2pt}}r@{\hspace{2pt}}
        @{\hspace{2pt}}r@{\hspace{0pt}}
      }
      \toprule
      LLMs & Method & \SCI & \Pantry & \Tools & \Toys & \INS\\
      \midrule
      \multirow{4}{*}{\DistillBERT}
      & \LoRA    & 228.1 & 621.6 & 1223.7 & 1644.2 & 667.6\\
      & \ADALoRA & 231.1 & 626.5 & 1245.0 & 1659.4 & 674.8\\
      & \method  &  \textbf{16.3} &  \textbf{28.3} &  \textbf{36.8} &   \textbf{25.7} &  \textbf{32.3}\\
      \midrule
      \multirow{4}{*}{\DistillRoBERTa}
      & \LoRA    & 229.7 & 604.1 & 1206.4 & 1702.0 & 673.6\\
      & \ADALoRA & 231.8 & 620.1 & 1231.9 & 1679.2 & 689.9\\
      & \method  &  \textbf{26.6} &  \textbf{26.6} &   \textbf{40.7} &   \textbf{29.4} &  \textbf{36.8}\\
      \midrule
      \multirow{4}{*}{\BERTMedium}
      & \LoRA    & 228.0 & 595.7 & 1143.0 & 1555.7 & 804.5\\
      & \ADALoRA & 238.8 & 626.5 & 1136.1 & 1743.7 & 805.8\\
      & \method  &  \textbf{11.8} &  \textbf{21.9} &  \textbf{30.6} &   \textbf{24.3} &  \textbf{26.2}\\
      \bottomrule
      \end{tabular}
      \begin{tablenotes}[normal,flushleft]
      \begin{footnotesize}
      \item
      The best run-time performance on each dataset is in \textbf{bold}.
      \par
      \end{footnotesize}
      %\end{scriptsize}
      \end{tablenotes}
  \vspace{-5pt}
  \end{threeparttable}
\end{table}

%\label{tbl:runtime}

Table~\ref{tbl:runtime} shows the run-time performance of \method, \LoRA and \ADALoRA in training.
As shown in Table~\ref{tbl:runtime}, \method substantially outperforms \LoRA and \ADALoRA in run-time efficiency during training.
Specifically, using \DistillBERT, \method achieves an average speedup of 30.8 and 31.1 compared to \LoRA and \ADALoRA, respectively.
A similar trend could also be observed when using \DistillRoBERTa and \BERTMedium.
Different from \LoRA and \ADALoRA which learn adapters inside LLMs, \method learns adapters separate from LLMs. 
As shown in Figure~\ref{fig:architecture}, this design allows \method to avoid both forward and backward propagation across LLMs in training, and thus, enable efficient adaptation.
We observe that \method, \LoRA and \ADALoRA require a similar number of epochs to converge.
For example, on \SCI, \method, \LoRA and \ADALoRA achieves the best validation \RT on the 77-th, 79-th and 62-th epoch, respectively, using \DistillBERT.
Thus, the run-time per epoch could serve as a valid measurement of the run-time efficiency in training for these methods.

%+++++++++++++++++++++++++++++++++++++++++++++++++++++++++++
\subsubsection{Comparison on Memory Efficiency}
\label{sec:results:efficiency:memory}
%+++++++++++++++++++++++++++++++++++++++++++++++++++++++++++

\begin{table}[!t]
\footnotesize
  \caption{Comparison on GPU Memory Usage (GB)}
  \centering
  \vspace{-10pt}
  \label{tbl:memory}
  \begin{threeparttable}
      \begin{tabular}{
        @{\hspace{0pt}}l@{\hspace{3pt}}
	  @{\hspace{3pt}}l@{\hspace{3pt}}
	  @{\hspace{3pt}}r@{\hspace{3pt}}
	  @{\hspace{3pt}}r@{\hspace{3pt}}
	  @{\hspace{3pt}}r@{\hspace{3pt}}
	  @{\hspace{3pt}}r@{\hspace{3pt}}
        @{\hspace{3pt}}r@{\hspace{0pt}}
      }
      \toprule
      LLMs & Method & \SCI & \Pantry & \Tools & \Toys & \INS\\
      \midrule
      \multirow{4}{*}{\DistillBERT}
      & \LoRA    & 21.0 & 15.1 & 30.8 & 76.9 & 50.1\\
      & \ADALoRA & 27.9 & 11.0 & 30.4 & 76.7 & 40.4\\
      & \method  &  \textbf{2.3} &  \textbf{1.5} &  \textbf{2.3} &  \textbf{1.8} &  \textbf{1.8}\\
      \midrule
      \multirow{4}{*}{\DistillRoBERTa}
      & \LoRA    & 22.6 & 12.7 & 30.8 & 77.5 & 41.2\\
      & \ADALoRA & 17.6 & 13.5 & 30.4 & 77.9 & 49.3\\
      & \method  &  \textbf{0.7} &  \textbf{1.5} &  \textbf{2.3} &  \textbf{2.6} &  \textbf{1.0}\\
      \midrule
      \multirow{4}{*}{\BERTMedium}
      & \LoRA    & 22.4 & 11.7 & 30.6 & 77.8 & 41.1\\
      & \ADALoRA & 19.7 & 11.0 & 30.5 & 77.5 & 37.2\\
      & \method  &  \textbf{1.3} &  \textbf{1.3} &  \textbf{2.1} &  \textbf{1.5} &  \textbf{1.5}\\
      \bottomrule
      \end{tabular}
      \begin{tablenotes}[normal,flushleft]
      \begin{footnotesize}
      \item
      In this table, the best performance on each dataset is in \textbf{bold}.
      \par
      \end{footnotesize}
      %\end{scriptsize}
      \end{tablenotes}
  \vspace{-10pt}
  \end{threeparttable}
\end{table}

%\label{tbl:memory}
%
Table~\ref{tbl:memory} presents the memory efficiency during training of \method, \LoRA and \ADALoRA on the five datasets when using different LLMs.
As shown in Table~\ref{tbl:memory}, \method achieves superior memory efficiency compared to \LoRA and \ADALoRA across all the datasets and LLMs.
For example, when training on \SCI using \DistillBERT, \method uses only 11.0\% of the GPU memory that \LoRA requires.
Similarly, on \Toys with \DistillBERT, \method uses a mere 2.3\% of the GPU memory required by both \LoRA and \ADALoRA, amounting to a 97.7\% reduction.
A similar trend could also be observed when using \DistillRoBERTa and \BERTMedium.

%**************************************************
\subsection{Analysis on Integration Strategies}
\label{sec:results:fusion}
%**************************************************

\begin{table}
\footnotesize
  \caption{Comparison on Integration Strategy (\DistillBERT)}
  \centering
  \vspace{-10pt}
  \label{tbl:analysis}
  \begin{threeparttable}
      \begin{tabular}{
        @{\hspace{0pt}}l@{\hspace{11pt}}
	  @{\hspace{11pt}}l@{\hspace{11pt}}
	  @{\hspace{11pt}}r@{\hspace{11pt}}
	  @{\hspace{11pt}}r@{\hspace{11pt}}
        @{\hspace{11pt}}r@{\hspace{0pt}}
      }
      \toprule
      Dataset & Metric & \MEAN & \WEIGHTED & \method\\
      \midrule
      \multirow{4}{*}{\SCI}
      & \RT & 0.0990 & 0.0976 & \textbf{0.1059}\\
      & \RF & 0.1815 & 0.1822 & \textbf{0.1986}\\
      \cline{2-5}
      & \NT & 0.0635 & 0.0616 & \textbf{0.0692}\\
      & \NF & 0.0816 & 0.0801 & \textbf{0.0892}\\
      \midrule
      \multirow{4}{*}{\Pantry}
      & \RT & 0.0526 & 0.0518 & \textbf{0.0581}\\
      & \RF & 0.1155 & 0.1106 & \textbf{0.1220}\\
      \cline{2-5}
      & \NT & 0.0286 & 0.0289 & \textbf{0.0323}\\
      & \NF & 0.0422 & 0.0418 & \textbf{0.0460}\\
      \midrule
      \multirow{4}{*}{\Tools}
      & \RT & 0.0593 & 0.0616 & \textbf{0.0673}\\
      & \RF & 0.1231 & 0.1241 & \textbf{0.1352}\\
      \cline{2-5}
      & \NT & 0.0413 & 0.0423 & \textbf{0.0460}\\
      & \NF & 0.0550 & 0.0558 & \textbf{0.0605}\\
      \midrule
      \multirow{4}{*}{\Toys}
      & \RT & 0.0787 & 0.0740 & \textbf{0.0855}\\
      & \RF & 0.1761 & 0.1755 & \textbf{0.1910}\\
      \cline{2-5}
      & \NT & 0.0446 & 0.0425 & \textbf{0.0499}\\
      & \NF & 0.0657 & 0.0644 & \textbf{0.0727}\\
      \midrule
      \multirow{4}{*}{\INS}
      & \RT & 0.0795 & 0.0781 & \textbf{0.0841}\\
      & \RF & 0.1422 & 0.1412 & \textbf{0.1516}\\
      \cline{2-5}
      & \NT & 0.0621 & 0.0611 & \textbf{0.0641}\\
      & \NF & 0.0757 & 0.0746 & \textbf{0.0786}\\
      \bottomrule
      \end{tabular}
      \begin{tablenotes}[normal,flushleft]
      \begin{footnotesize}
      \item
      In this table, the best performance on each dataset is in \textbf{bold}.
      \par
      \end{footnotesize}
      %\end{scriptsize}
      \end{tablenotes}
  \vspace{-10pt}
  \end{threeparttable}
\end{table}
%\label{tbl:analysis}

We conduct an analysis to investigate the effectiveness of integrating adapters sequentially.
To this end, we introduce two \method variants \MEAN and \WEIGHTED.
\MEAN and \WEIGHTED differ from \method in the ways of integrating adapters.
Specifically, \MEAN aggregates the outputs of adapters using a mean pooling.
\WEIGHTED learns weights on different adapters, and aggregates the outputs of adapters using a weighted sum.
We compare \method against \MEAN and \WEIGHTED to evaluate the effectiveness of integrating adapters sequentially.
Table~\ref{tbl:analysis} shows the recommendation performance of \method, \MEAN and \WEIGHTED on the five datasets using \DistillBERT.
Due to the space limit, 
we only present the results from \DistillBERT. 
But we observe a similar trend when using \DistillRoBERTa and \BERTMedium.

As presented in Table~\ref{tbl:analysis}, \method outperforms both \MEAN and \WEIGHTED on all the five datasets.
Particularly, in terms of \RT, 
\method achieves a remarkable average improvement of 9.7\% and 11.3\% compared to \MEAN and \WEIGHTED, respectively, over the five datasets.
Similarly, in terms of \NT, \method substantially outperforms \MEAN and \WEIGHTED with an average improvement of 10.4\% and 11.7\%, respectively, across the five datasets.
These results show the effectiveness of integrating adapters in a sequential manner in \method.

%**************************************************
\subsection{Parameter Study}
\label{sec:results:para}
%**************************************************

\begin{figure}[!h]
	%\vspace{-5pt}
	\centering
	\footnotesize
	\begin{subfigure}{0.48\linewidth}
		\centering
		%\hspace*{20pt}
		\includegraphics[width=\linewidth]{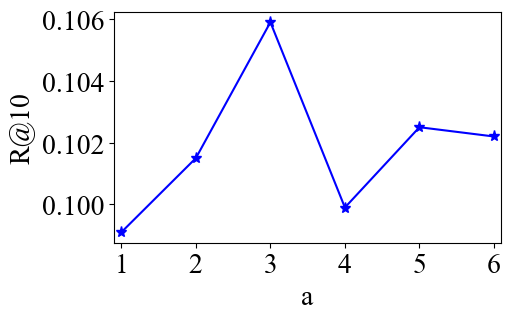}
		%\vspace*{10pt}
		\caption{\SCI}
		\label{fig:para:sci}
	\end{subfigure}
	\begin{subfigure}{0.45\linewidth}
		\centering
		%\hspace*{20pt}
		\includegraphics[width=\linewidth]{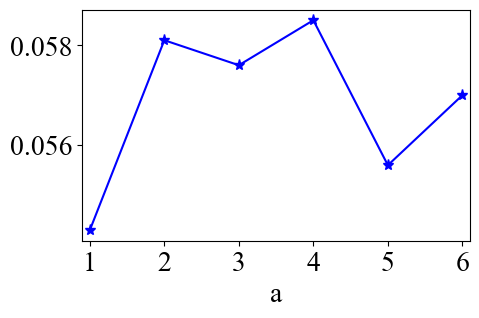}
		\caption{\Pantry}
		\label{fig:para:pantry}
	\end{subfigure}
 \\
	\begin{subfigure}{0.48\linewidth}
		\centering
		%\hspace*{20pt}
		\includegraphics[width=\linewidth]{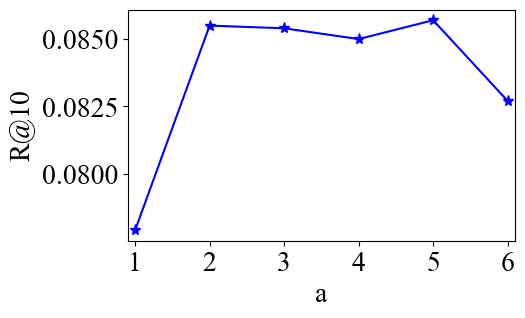}
		\caption{\Toys}
		\label{fig:para:toys}
	\end{subfigure}
	\begin{subfigure}{0.44\linewidth}
		\centering
		%\hspace*{20pt}                    
		\includegraphics[width=\linewidth]{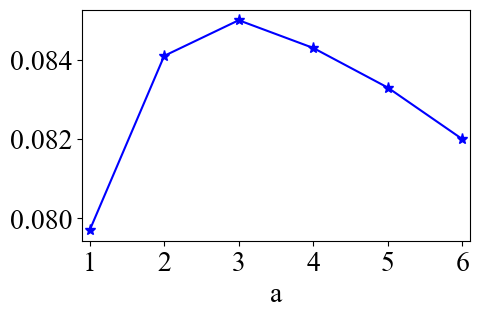}                  
		\caption{\INS}
		\label{fig:para:ins}
	\end{subfigure}
%
%\vspace{-10pt}
\caption{Recommendation performance of \method over the number of layers to adapt (i.e., $a$)} 
%\xia{this is on 
%which method? align the figures}}
\label{fig:para}
\vspace{-10pt}
\end{figure}

We conduct a parameter study to evaluate how recommendation performance changes over the number of top LLM layers to be adapted in \method (i.e., $a$).
Particularly, we utilize the widely used \SCI, \Pantry, \Toys and \INS datasets for this study. %\xia{why not all the datasets?}.
On each dataset, we change $a$ 
%\xia{need to remind what $a$ is here} 
while fixing all the other hyper-parameters as the ones reported in Table~\ref{tbl:para_distillbert} (Appendix).
Figure~\ref{fig:para} shows the performance at \RT on the four datasets.

As shown in Figure~\ref{fig:para}, on all the four datasets, adapting multiple layers ($a\geq2$) consistently outperforms that of adapting only the top layer of LLMs ($a=1$).
These results demonstrate the effectiveness of jointly adapting the top-$a$ layers ($a\geq2$) in \method.
It is also worth noting that, 
as shown in Figure~\ref{fig:para}, 
adapting all the LLM layers ($a=6$) may not benefit the recommendation performance.
In fact, on all the datasets, \method achieves the best performance when $a<6$.
%\xia{you need to provide insights as to why only a few layers (what those layers learn and represent?)}
%
A possible reason could be that recommendation datasets are generally sparse.
Adapting all the LLM layers on sparse recommendation datasets could induce overfitting, and thus, degrade the recommendation performance.

%%%%%%%%%%%%%%%%%%%%%%%%%%%%%%%%%%%%%%%%%%%%%%%%%
\section{Conclusion}
\label{sec:conclusion}
%%%%%%%%%%%%%%%%%%%%%%%%%%%%%%%%%%%%%%%%%%%%%%%%%

In this paper, we present a novel LLM adaptation method \method for LLM-enhanced SR.
%
%\method allows both efficient and effective LLM adaptation by 
%1) learning adapters separate from LLMs while fixing all the parameters within LLMs; 
%2) jointly adapting the top-$a$ layers of LLMs; and 3) integrating the adapters in a sequential manner.
%
\method allows efficient adaptation by learning adapters separate from LLMs and fixing all the pre-trained parameters within LLMs to avoid forward and backward propagation across LLMs in training.
\method also enables effective adaptation by jointly adapting the top-$a$ layers of LLMs, 
and integrating adapters sequentially.
We evaluate \method against five state-of-the-art baseline methods on five benchmark datasets using three different LLMs.
The experimental results demonstrate that \method substantially outperforms all the baseline methods in terms of recommendation performance, and achieves remarkable improvement in terms of run-time and memory efficiency in training when compared to the best-performing baseline methods \LoRA and \ADALoRA (e.g., 30.8x average speedup over \LoRA across the five datasets).
Our analysis also shows the effectiveness of integrating adapters sequentially in \method.
Our parameter study suggests that adapting top-$a$ layers ($a \geq 2$) of LLMs together in \method substantially benefits the recommendation performance.
Our parameters study also shows that adapting all LLM layers for SR tasks may not benefit the recommendation performance.

\clearpage

% Use \bibliography{yourbibfile} instead or the References section will not appear in your paper
\bibliography{main}

\clearpage

%%%%%%%%%%%%%%%%%%%%%%%%%%%%%%%%%%%%%%%%%%%%%
\appendix
%%%%%%%%%%%%%%%%%%%%%%%%%%%%%%%%%%%%%%%%%%%%%
\section*{Appendix}

\setcounter{secnumdepth}{2} %May be changed to 1 or 2 if section numbers are desired.

\setcounter{section}{0}
\renewcommand{\thesection}{S\arabic{section}}

\setcounter{table}{0}
\renewcommand{\thetable}{S\arabic{table}}

\setcounter{figure}{0}
\renewcommand{\thefigure}{S\arabic{figure}}

\setcounter{algorithm}{0}
\renewcommand{\thealgorithm}{S\arabic{algorithm}}

%%%%%%%%%%%%%%%%%%%%%%%%%%%%%%%%%%%%%%%%%%%%%
\section{Best-performing Hyper-parameters}
\label{supp:parameter}
%%%%%%%%%%%%%%%%%%%%%%%%%%%%%%%%%%%%%%%%%%%%%

Table~\ref{tbl:para_distillbert}, Table~\ref{tbl:para_distillroberta} and Table~\ref{tbl:para_bertmedium} shows the best-performing hyper-parameters of \method and all the baseline methods when using \DistillBERT, \DistillRoBERTa and \BERTMedium, respectively.

\begin{table}[!h]
\footnotesize
  \caption{Best-performing Hyper-parameters (\DistillBERT)}
  \centering
  \vspace{-10pt}
  \label{tbl:para_distillbert}
  \begin{threeparttable}
      \begin{tabular}{
        @{\hspace{0pt}}l@{\hspace{5pt}}
	  @{\hspace{5pt}}r@{\hspace{5pt}}
        @{\hspace{1pt}}r@{\hspace{1pt}}
	  @{\hspace{5pt}}r@{\hspace{5pt}}
        @{\hspace{1pt}}r@{\hspace{1pt}}
	  @{\hspace{5pt}}r@{\hspace{5pt}}
        @{\hspace{1pt}}r@{\hspace{1pt}}
	  @{\hspace{5pt}}r@{\hspace{5pt}}
	  @{\hspace{5pt}}r@{\hspace{5pt}}
        @{\hspace{1pt}}r@{\hspace{1pt}}
        @{\hspace{5pt}}r@{\hspace{5pt}}
        @{\hspace{1pt}}r@{\hspace{1pt}}
        @{\hspace{5pt}}r@{\hspace{5pt}}
        @{\hspace{5pt}}r@{\hspace{0pt}}
      }
      \toprule
      \multirow{2}{*}{Dataset} 
      & \LoRA && \ADALoRA && \Full && \multicolumn{2}{c}{\TTMLP} 
      && \TTMoE && \multicolumn{2}{c}{\method}\\
      \cmidrule{2-2} \cmidrule{4-4} \cmidrule{6-6} \cmidrule{8-9} \cmidrule{11-11} \cmidrule{13-14}
      & $p$ && $p$ && $n_t$ && $b$ & $n_l$ && $b$ && $b$ & $a$\\
      \midrule
      \SCI    & 0.1 && 0.2 && 2 && 256 & 3 && 128 && 256 & 3\\
      \midrule
      \Pantry & 0.3 && 0.1 && 2 && 256 & 2 && 128 && 256 & 2\\
      \midrule
      \Tools  & 0.1 && 0.2 && 2 && 256 & 3 && 256 && 256 & 3\\
      \midrule
      \Toys   & 0.3 && 0.3 && 2 && 256 & 3 && 256 && 256 & 2\\
      \midrule
      \INS    & 0.3 && 0.3 && 4 && 256 & 2 && 256 && 256 & 2\\
      \bottomrule
      \end{tabular}
      \begin{tablenotes}[normal,flushleft]
      \begin{footnotesize}
      \item
      This table shows the best-performing hyper-parameters of \method and all the baseline methods on the five datasets when using \DistillBERT.
      \par
      \end{footnotesize}
      %\end{scriptsize}
      \end{tablenotes}
  \vspace{-5pt}
  \end{threeparttable}
\end{table}
%\label{tbl:para_distillbert}

\begin{table}[!h]
\footnotesize
  \caption{Best-performing Hyper-parameters (\DistillRoBERTa)}
  \centering
  \vspace{-10pt}
  \label{tbl:para_distillroberta}
  \begin{threeparttable}
      \begin{tabular}{
        @{\hspace{0pt}}l@{\hspace{5pt}}
	  @{\hspace{5pt}}r@{\hspace{5pt}}
        @{\hspace{1pt}}r@{\hspace{1pt}}
	  @{\hspace{5pt}}r@{\hspace{5pt}}
        @{\hspace{1pt}}r@{\hspace{1pt}}
	  @{\hspace{5pt}}r@{\hspace{5pt}}
        @{\hspace{1pt}}r@{\hspace{1pt}}
	  @{\hspace{5pt}}r@{\hspace{5pt}}
	  @{\hspace{5pt}}r@{\hspace{5pt}}
        @{\hspace{1pt}}r@{\hspace{1pt}}
        @{\hspace{5pt}}r@{\hspace{5pt}}
        @{\hspace{1pt}}r@{\hspace{1pt}}
        @{\hspace{5pt}}r@{\hspace{5pt}}
        @{\hspace{5pt}}r@{\hspace{0pt}}
      }
      \toprule
      \multirow{2}{*}{Dataset} 
      & \LoRA && \ADALoRA && \Full && \multicolumn{2}{c}{\TTMLP} 
      && \TTMoE && \multicolumn{2}{c}{\method}\\
      \cmidrule{2-2} \cmidrule{4-4} \cmidrule{6-6} \cmidrule{8-9} \cmidrule{11-11} \cmidrule{13-14}
      & $p$ && $p$ && $n_t$ && $b$ & $n_l$ && $b$ && $b$ & $a$\\
      \midrule
      \SCI    & 0.2 && 0.3 && 2 && 256 & 3 && 128 &&  64 & 3\\
      \midrule
      \Pantry & 0.3 && 0.2 && 2 && 256 & 1 && 256 && 256 & 2\\
      \midrule
      \Tools  & 0.1 && 0.2 && 2 && 128 & 3 &&  64 && 256 & 3\\
      \midrule
      \Toys   & 0.3 && 0.3 && 2 && 256 & 3 && 256 && 256 & 2\\
      \midrule
      \INS    & 0.3 && 0.3 && 2 && 256 & 2 && 256 && 256 & 2\\
      \bottomrule
      \end{tabular}
      \begin{tablenotes}[normal,flushleft]
      \begin{footnotesize}
      \item
      This table shows the best-performing hyper-parameters of \method and all the baseline methods on the five datasets when using \DistillRoBERTa.
      \par
      \end{footnotesize}
      %\end{scriptsize}
      \end{tablenotes}
  \vspace{-5pt}
  \end{threeparttable}
\end{table}
%\label{tbl:para_distillroberta}

\begin{table}[!h]
\footnotesize
  \caption{Best-performing Hyper-parameters (\BERTMedium)}
  \centering
  \vspace{-10pt}
  \label{tbl:para_bertmedium}
  \begin{threeparttable}
      \begin{tabular}{
        @{\hspace{0pt}}l@{\hspace{5pt}}
	  @{\hspace{5pt}}r@{\hspace{5pt}}
        @{\hspace{1pt}}r@{\hspace{1pt}}
	  @{\hspace{5pt}}r@{\hspace{5pt}}
        @{\hspace{1pt}}r@{\hspace{1pt}}
	  @{\hspace{5pt}}r@{\hspace{5pt}}
        @{\hspace{1pt}}r@{\hspace{1pt}}
	  @{\hspace{5pt}}r@{\hspace{5pt}}
	  @{\hspace{5pt}}r@{\hspace{5pt}}
        @{\hspace{1pt}}r@{\hspace{1pt}}
        @{\hspace{5pt}}r@{\hspace{5pt}}
        @{\hspace{1pt}}r@{\hspace{1pt}}
        @{\hspace{5pt}}r@{\hspace{5pt}}
        @{\hspace{5pt}}r@{\hspace{0pt}}
      }
      \toprule
      \multirow{2}{*}{Dataset} 
      & \LoRA && \ADALoRA && \Full && \multicolumn{2}{c}{\TTMLP} 
      && \TTMoE && \multicolumn{2}{c}{\method}\\
      \cmidrule{2-2} \cmidrule{4-4} \cmidrule{6-6} \cmidrule{8-9} \cmidrule{11-11} \cmidrule{13-14}
      & $p$ && $p$ && $n_t$ && $b$ & $n_l$ && $b$ && $b$ & $a$\\
      \midrule
      \SCI    & 0.2 && 0.1 && 2 && 256 & 3 &&  64 && 256 & 2\\
      \midrule
      \Pantry & 0.2 && 0.1 && 2 && 256 & 3 && 256 && 256 & 2\\
      \midrule
      \Tools  & 0.2 && 0.1 && 2 && 256 & 3 &&  64 && 256 & 3\\
      \midrule
      \Toys   & 0.3 && 0.3 && 2 && 256 & 3 && 256 && 256 & 2\\
      \midrule
      \INS    & 0.3 && 0.3 && 4 && 256 & 3 && 256 && 256 & 2\\
      \bottomrule
      \end{tabular}
      \begin{tablenotes}[normal,flushleft]
      \begin{footnotesize}
      \item
      This table shows the best-performing hyper-parameters of \method and all the baseline methods on the five datasets when using \BERTMedium.
      \par
      \end{footnotesize}
      %\end{scriptsize}
      \end{tablenotes}
  \vspace{-5pt}
  \end{threeparttable}
\end{table}
%\label{tbl:para_bertmedium}

\newpage

\end{document}